\documentclass{article}

\usepackage{arxiv}

\usepackage[utf8]{inputenc} % allow utf-8 input
\usepackage[T1]{fontenc}    % use 8-bit T1 fonts
\usepackage{hyperref}       % hyperlinks
\usepackage{url}            % simple URL typesetting
\usepackage{booktabs}       % professional-quality tables
\usepackage{amsfonts}       % blackboard math symbols
\usepackage{nicefrac}       % compact symbols for 1/2, etc.
\usepackage{microtype}      % microtypography
\usepackage{lipsum}
\usepackage{graphicx}
\usepackage{algorithm}
\usepackage{algpseudocode}
\usepackage{tabularx}
\usepackage{adjustbox}
\usepackage{amsmath}    
\graphicspath{ {./images/} }

\title{Borhan: A Novel System for Prioritized Default Logic}

\author{
 Alireza Shahbazi \\
  School of Computer Engineering\\
  Iran University of Science and Technology\\
  Tehran, Iran \\
  \texttt{shahbazi@borhan-onto.ir} \\
   \And
 Mohammad Hossein Khojasteh \\
  School of Computer Engineering\\
  Iran University of Science and Technology\\
  Tehran, Iran \\
  \texttt{m\_khojaste@comp.iust.ac.ir} \\
  \And
 Behrouz Minaei-Bidgoli \\
  School of Computer Engineering\\
  Iran University of Science and Technology\\
  Tehran, Iran \\
  \texttt{b\_minaei@iust.ac.ir} \\
}

\begin{document}
\maketitle
\begin{abstract}
     Prioritized Default Logic presents an optimal solution for addressing real-world problems characterized by incomplete information and the need to establish preferences among diverse scenarios. Although it has reached great success in the theoretical aspect, its practical implementation has received less attention. In this article, we introduce Borhan, a system designed and created for prioritized default logic reasoning. To create an effective system, we have refined existing default logic definitions, including the extension concept, and introduced novel concepts. In addition to its theoretical merits, Borhan proves its practical utility by efficiently addressing a range of prioritized default logic problems. In addition, one of the advantages of our system is its ability to both store and report the explanation path for any inferred triple, enhancing transparency and interpretability. Borhan is offered as an open-source system, implemented in Python, and even offers a simplified Java version as a plugin for the Protege ontology editor. Borhan thus represents a significant step forward in bridging the gap between the theoretical foundations of default logic and its real-world applications.
  \end{abstract}

% keywords can be removed

\section{Introduction}
The defeasible inference is a type of inference that is suitable for problems that suffer from a lack of information, and the conclusions of reasoners are based on current knowledge. They can be retracted in light of newly available information. A family of formal frameworks formulated to represent defeasible inference is called non-monotonic logic \cite{sep-logic-nonmonotonic}. The challenge of non-monotonic reasoning has been a long-standing problem, captivating the attention of numerous researchers who have sought to expand its logical foundations or develop reasoning systems. In their seminal 1980 article on non-monotonic reasoning, McDermott et al. stated a fundamental distinction between this form of reasoning and classical logic: in non-monotonic systems, the introduction of a new axiom can render a previously established axiom invalid \cite{mcdermott1980non}.

Default logic, a widely adopted method for non-monotonic reasoning, has attracted great attention and research in recent years. This popularity can be attributed to the simplicity of the default concept and its widespread applicability in various domains \cite{reiter1980logic, garcia2004defeasible, bozzato2022reasoning}. While classical reasoning suffices for solving problems when complete information is accessible, the true challenge arises in situations where information is incomplete or there's insufficient time to gather all necessary data. To illustrate, imagine an emergency scenario where a doctor must make an initial treatment decision based on the most likely causes due to a lack of time to wait for all test results.

The main focus of this article is on prioritized default logic \cite{brewka1994reasoning}. Sometimes there are situations when more than one default can be applied. In these situations, there should be priorities between defaults to show preferences on which default to apply. For example consider the default theory $T = (W, D)$ where $W = \{bird, penguin\}$ and $D$ indicating following defaults: \par
\[ \delta_1 = \frac{penguin : \neg flies}{\neg flies}, \delta_2 = \frac{bird:flies}{flies} \] 
In this case, $\delta_1$ is preferred over $\delta_2$, thus $T$ should only admit the extension $Th(\{bird, penguin, \neg flies\})$. \par

The absence of a comprehensive system capable of addressing various problems in prioritized default logic motivated the development of a new system. Initially, we identified problems in prioritized default logic where existing definitions proved inadequate to solve them. To overcome this challenge, we introduced a novel definition for the extension concept, effectively resolving the previously mentioned issues. Following the refinement of theoretical concepts, we devised an innovative system architecture. Borhan is a dual-component system comprising a descriptive and a default logic reasoner component. The descriptive logic reasoner's primary function is to assess consistency, invoking the default logic reasoner component in the presence of inconsistencies. The default logic reasoner component, in turn, handles these inconsistencies and executes default rules to derive new triples. The process of adding new triples and checking for inconsistencies continues until the graph model reaches a fixed point, remaining unchanged thereafter.

To evaluate the efficacy of our proposed model, we assembled various domain-specific examples, including problems that previous definitions and systems failed to address. In contrast to other models, our approach successfully handles all these examples.

Our system's contributions can be summarized as follows:

\begin{enumerate}
    \item Introduction of a novel definition for the extension concept, enabling to handling of new problems.
    \item Development of an innovative and explainable system capable of addressing a wide range of issues in prioritized default logic.
\end{enumerate}

The rest of the paper is organized as follows. First, we commence with an exploration of related works in non-monotonic and default reasoning to establish context. Subsequently, we delve into a comprehensive exposition of our proposed model, explaining its details. Finally, we conclude with a comparative analysis, evaluating our system in contrast to its predecessors, and highlighting the advancements and contributions of our approach.

\section{Related Work}
In this section, we first introduce non-monotonic reasoning, providing an initial glimpse into its fundamental concept. We then briefly explore projects that have extended this logic and designed systems around it. Following this, we delve into the logical articles with the subject of default reasoning as one of the formations of non-monotonic reasoning. Lastly, we will explain deep learning models tailored for non-monotonic reasoning, introducing the use of modern neural networks to address this problem. \par

\subsection{Non-Monotonic Reasoning}
As previously mentioned, non-monotonic reasoning has presented a long-standing challenge. Consider, for instance, the axiom "the cup has dropped," which initially leads to the conclusion that the cup is broken. However, the incorporation of new information, such as the cup being made of metal or landing on a pile of soft pillows, can challenge the accuracy of the initial inference \cite{rudinger2020thinking}. The significance of non-monotonic reasoning arises from the recognition that our knowledge is often incomplete, necessitating the ability to revise or remove prior knowledge in light of new insights. To address this realm of reasoning, McDermott et al. have developed various formal frameworks, including Proof Theories, Model Theory, and Fixed-point Theories, tailored specifically for non-monotonic reasoning. These frameworks provide a structured foundation for reasoning in situations where traditional, monotonic logic falls short \cite{mcdermott1980non}. \par 

Various formalisms exist for non-monotonic reasoning, with each offering distinct approaches to address the challenges posed by incomplete or evolving knowledge. These formalisms encompass the closed-world assumption, argument-based methods, default logic, autoepistemic logic, selection semantics, and assumption-based techniques \cite{sep-logic-nonmonotonic}. Each of these formalisms has garnered considerable attention in research, with efforts aimed at their expansion and the creation of corresponding systems. As an illustrative example, a recent study conducted by Arieli et al. delves into the realm of argument-based approaches, one of these formalisms. In their comprehensive investigation, the authors conducted an abstract study of logical argumentation frameworks. Their study involved a classification of these frameworks, coupled with an assessment of the desiderata satisfied by each class. This research contributes to our understanding of the capabilities and limitations of argument-based approaches within the domain of non-monotonic reasoning \cite{arieli2023postulate}. \par

Various systems have been developed to handle non-monotonic reasoning, expanding the capabilities of logic-based programming languages. Among these, Prolog, a well-known logic programming language traditionally used for classical logic reasoning, has proposed an extension to address non-monotonic reasoning by Chen et al \cite{chen1996extending}. Another noteworthy example is Disjunctive Logic Programming (DLP), a logic programming that permits disjunctions in the heads of rules, causing more expressive language than disjunction-free logic programming. DLV stands out as a system that effectively implements disjunctive logic programming, capable of handling a wide range of logic programming tasks, including those related to non-monotonic reasoning. In addition to DLV, DLV+ represents a notable advancement \cite{leone2006dlv}. DLV+ introduces object-oriented constructs, enhancing its capabilities as an improvement over the original DLV system \cite{ricca2007disjunctive}. These systems collectively exemplify the ongoing evolution of non-monotonic reasoning, demonstrating the adaptation and integration of flexible reasoning paradigms within the realm of logic programming.

\subsection{Default Reasoning}
The axiom "A bird can fly" is universally accepted as true. However, exceptions exist, such as penguins. To express the correct axiom, we should specify, "A bird can fly as long as it is not a penguin." This form of reasoning, where something is considered true by default unless exceptions are present, is known as default reasoning, initially proposed by Raymond Reiter. Various variations of default reasoning exist, including weak extension default logic, disjunctive default logic, and prioritized default logic, with our focus in this article being on the latter \cite{reiter1980logic}. \par 

A variety of programming languages and systems have been proposed to implement Default Logic reasoner, each offering unique approaches and features. DeLP (Defeasible Logic Programming) combines the principles of Logic Programming and Defeasible Argumentation. Within this programming language, queries can be countered by opposing arguments. DeLP also includes considerations for default negation \cite{garcia2004defeasible}. DeReS, developed by Truszczy´nski et al., is an automated reasoning system tailored for the implementation of Reiter's default logic. It facilitates fundamental reasoning tasks, including the identification of all extensions \cite{cholewinski1996default}. A default logic solver named dl2asp employs a translation approach, converting default logic into answer set programming (ASP) by unveiling internal relationships between formulas in a default theory \cite{chen2010dl2asp}. Bozzato et al. introduced a framework for characterizing OWL RL knowledge bases, integrating the concept of justifiable exceptions related to defensible axioms. Reasoning within this framework is achieved through a translation into ASP programming, with a specific focus on the limited version of $DL-Lite_R$ axioms, simplifying ASP encoding \cite{bozzato2022reasoning}. Logic Programming with Ordered Disjunction (LPOD) aims to address preference handling by introducing the concept of $A \times B$ to establish the priority between head rules. This definition signifies that option A is the most preferred choice. If A is valid, it should be considered the definitive answer. However, if A is not valid, then B represents the correct, or at least a valid, definitive answer \cite{brewka2002logic}. \par

These programming languages and systems collectively contribute to the rich landscape of Default Logic, offering solutions for various aspects of non-monotonic reasoning and preference modeling.

\subsection{Neural Models for Non-Monotonic Reasoning}
In recent years, significant advancements in neural models have opened up new possibilities for tackling diverse tasks. Among these, Large Language Models (LLMs) represent a notable breakthrough in the realm of neural models, enabling the attainment of high accuracy across various natural language processing tasks \cite{devlin2018bert}. \par 

One such task is Natural Language Inference (NLI), where the primary objective is to assess whether a hypothesis can be logically inferred from a set of premises \cite{yu2023nature}. In pursuit of advancing the field of defeasible inference, Rudinger et al. have introduced a novel dataset known as "Defeasible NLI." This dataset comprises rows, each consisting of three distinct sentences: a premise, a hypothesis, and an update. The dataset's purpose is to train models capable of determining whether the update sentence strengthens or weakens the hypothesis based on the provided premise \cite{rudinger2020thinking}. \par 

For instance, given a premise such as "Two men and a dog are standing among rolling green hills" and a hypothesis like "The men are farmers," the introduction of a new update such as "The men are wearing backpacks" weakens the hypothesis, while "The dog is a sheepdog" strengthens it. Following the creation of this dataset, Rudinger et al. conducted experiments employing LLMs such as RoBERTa \cite{liu2019roberta}, T5 \cite{raffel2020exploring}, Bart \cite{lewis2019bart}, and GPT \cite{radford2019language}. \par 

The results of the experiments have shown that Large Language Models (LLMs) can achieve remarkably high levels of accuracy, nearly matching human-level performance in the Defeasible NLI task. However, a major limitation of these models is their inability to generate explanations. This lack of explanatory capability makes them unreliable and unsuitable for use in high-stakes domains, such as medicine or legal judgment. Nevertheless, these findings represent a substantial advancement in the utilization of neural models for defeasible inference and natural language comprehension.

\section{Problem Definition}
In this section, we will utilize Description Logic, the primary Classical Logic (CL) in our implementation, to provide a comprehensive explanation of Default Logic. The subsequent section will introduce the components of the Borhan architecture using these symbols and principles. To achieve this, we will draw upon the definitions and theorems from \cite{baader2003description} and \cite{baader1995embedding}.

\subsection{Default Theory}
A default theory is an ordered pair and can be considered as $T = (W, D)$ in which $W$ is the facts and axioms that are stated in Description Logic (including Tbox, Rbox, Abox), and $D$ consists of a set of default rules.

In the context of default theories, they are often represented as ordered pairs, typically denoted as $T = (W, D)$. Here, $W$ encompasses the collection of facts and axioms stated within Description Logic, including Tbox, Rbox, and Abox. Complementing this, $D$ constitutes a set of default rules, a crucial aspect of the default theory's structure. This combination of facts, axioms, and default rules forms the foundation for reasoning and inference within the framework of default logic. Within the Borhan framework, $W$ constitutes an OWL-DL model incorporating class, property, and individual definitions and their relations. This model can be stored in various formats like RDF-XML or TTL. Also, $D$ contains default rules that are in the default logic structure, enabling non-monotonic reasoning capabilities.

\subsection{Consequences in Classical Logic}
In the modeling of knowledge in the structure of Descriptive Logic, some parts of the knowledge are explicitly expressed, which are interpreted as Assertions. Despite Assertions, there is another part that is not explicitly stated, but it can be inferred by rules in Classical Logic. This inference can be made in both the Tbox and Abox sections. For Tbox's example,  suppose you know that every father is a man and every man is a human. In this case, it is not explicitly stated that every father is a human, but it is part of the facts of the world. Also, as an example for Abox, we may know that person P1 is the brother of person P2 and person P3 is P1's son, and also there is a logical rule which indicates that every man's father's brother is his uncle. In this case, although it is not explicitly stated that P2 is P3's uncle, this fact can be inferred from the given knowledge. The symbol $Th(W)$ is used to refer to all the facts of the world, whether they are explicitly given (Assertion) or hidden and requiring inference (Inferred). \par

In the knowledge modeling structure of Description Logic, knowledge is partitioned into two categories: explicit expressions known as Assertions and implicit inferences derived through Classical Logic rules. These inferences can be drawn from both the Tbox and Abox sections. For example, in the Tbox, if we know that every father is a man and every man is a human, it's not explicitly stated that every father is a human, but it is an implicit fact in the world's knowledge. Likewise, in the Abox, if we know that person P1 is the brother of person P2, and person P3 is P1's son, and there's a logical rule stating that every man's father's brother is his uncle, we can infer that P2 is P3's uncle, even if it's not explicitly mentioned. To refer to the entirety of world knowledge, whether explicitly given (Assertion) or requiring inference (Inferred), we use the symbol $Th(W)$.

\subsection{Default Rule}
In Classical Logic, rules consist of only two parts – premise and conclusion. In the context of default rules, however, they encompass three primary components. Moreover, in default rules with priority, an additional element is incorporated to manage preferences. An example of a default rule is provided below, which different parts will be explained in the following.

\begin{equation}
D_i = \frac{P_i:J_1, J_2, ..., J_n}{C_i}
\end{equation}

\textbf{Prerequisite:} The central role in the execution of a default rule is played by its prerequisites, which can be a well-structured combination of atomic statements. Each default rule's prerequisites must be satisfied for the rule's execution. The term "prerequisite" is used instead of "premise," as in Classical Logic, because it alone does not constitute a sufficient condition for the conclusion.

\textbf{Justification:} Default rules are executed only when their justifications align with our current knowledge, despite the prerequisites being met. Each rule can have multiple justifications, and how we interact with these justifications gives rise to various variants of Default Logic.

\textbf{Conclusion: } When the prerequisites of a rule are satisfied, and the rule's justifications are compatible with our existing knowledge, the result of that rule is incorporated into the current knowledge.

\textbf{Order: } In prioritized default logic, the conclusions added into knowledge via default rules are explicitly assigned rankings. This prioritization offers a significant advantage by enabling the resolution of inconsistencies when they arise. Specifically, if conflicts emerge among default knowledge, maintaining consistency becomes feasible by discarding the less valuable information. These rankings can be static and predefined during rule definition or can dynamically adapt to different situations or over time, as discussed in \cite{brewka1994reasoning}. The notation $Di >> Dj$ denotes that default rule $i$ holds greater value in comparison to default rule $j$.

A closed default rule is characterized by the absence of free variables within its prerequisites, justifications, and conclusions. Consequently, a default theory is considered closed when all of its rules are closed. Conversely, a default rule containing free variables is termed an open default rule.

Within the Borhan framework, rule definitions utilize SPIN (SPARQL Inferencing Notation). The prerequisites of the rule are placed within the WHERE body of the SPARQL command, the negation of justifications is accommodated in the FILTER NOT EXISTS section within the WHERE part, and the conclusions are placed in the CONSTRUCT section of the command.

For example, consider the common default rule "birds usually fly." This rule is expressed in the syntax of default logic as follows:

\begin{equation}
D_0 = \frac{Bird(x):Fly(x)}{Fly(x)}
\end{equation}
In the structure of SPIN rules of the Borhan framework, this rule will be expressed as follows: \par
$CONSTRUCT \{ ?x \; a \; :\!Fly.\}$ \par
$WHERE \{ ?x \; a \; :\!Bird.$ \par
$FILTER \; NOT \; EXISTS \{ ?x \;a \;?c1 .$ \par
$\;\; ?c1\; a \;owl:Class ; $ \par
$\;\;\;\; owl:complementOf \; :\!Fly.\} \}$

Every rule within the framework is assigned a predefined order, contained within the SPIN command alongside its name and comment. This predetermined order is subsequently incorporated as a Reification for all results derived from the corresponding SPARQL command. In our framework, this order is represented as a numeric value during rule definition. In a manner akin to order, the justifications for each rule are similarly preserved as Reifications for the results generated by that rule. Consequently, the results, including their order values and justifications, are seamlessly integrated into our basic knowledge.

Within this framework, the order values are assumed to be static and predetermined. Two underlying reasons support this assumption: a) in the majority of industrial scenarios, assigning rule priorities is feasible, and b) accommodating dynamic order modeling introduces complexity into the framework, as discussed in \cite{antoniou1999tutorial}. It's important to clarify that, akin to the approach presented in \cite{baader1995embedding}, the Borhan framework exclusively addresses default rules within the terminological structure. For instance, it does not support default inference for establishing subclass relationships between classes by default.

\subsection{Fixed Point}
In simple terms, a fixed point in Default Logic signifies a consistently expanded state from the current knowledge, where no further default rules can be executed. It's worth noting that not every default theory necessarily possesses a fixed point, as this characteristic is inherent to the nature of default rules. For instance, consider the rule: "Doctors usually have a child who is also a doctor." In this scenario, the fact that every doctor has a child who is a doctor by default, and that child, in turn, has a child who is a doctor by default due to their profession, and so on in an infinite loop, is a part of the logical framework's semantics. It's not necessarily an undesirable phenomenon, but it can introduce complexities in its practical implementation.

To address these challenges, we can draw inspiration from the approach presented in \cite{baader1995embedding}, which employed restricted semantics to solve issues arising from Skolemization. One potential solution is to limit the implementation of default rules exclusively to the individuals explicitly introduced in the knowledge base. Another approach involves identifying the group of rules prone to infinite calculations and implementing a mechanism to halt these calculations. For instance, in the earlier example, upon encountering such rules and generating multiple conclusions, the process of producing further conclusions can be terminated, with a report indicating that the process would otherwise continue indefinitely.

\subsection{Extension}
The concept of "extension" is fundamental in default logic, denoting the attainment of a consistent expansion of the basic knowledge base through the execution of default rules. To ensure this consistency in knowledge extension, it is essential to examine Classical Logic for consistency both before and after the execution of default rules. It's important to note that a theoretical default may yield various extensions, or in some cases, no extension at all. In our implementation, we have utilized the idea of prioritized default logic, wherein the prioritization of default rules typically results in at least one extension, and, in most instances, these priorities lead to a singular extension \cite{sep-logic-nonmonotonic}.

When dealing with multiple extensions, we can adopt a Skeptical approach, wherein we regard only the shared elements among these extensions as the ultimate theory result. However, in the framework of Prioritized Default Logic, we typically aim to establish a single extension through the prioritization of default rules. Alternatively, we may opt for a Credulous approach, treating each extension as a distinct expansion of the basic knowledge.

The normal extension ($E$) for a Default Theory $T = (W, D)$ is the smallest consistent extension of the basic knowledge that has the following properties:

\begin{itemize}
  \item Any extension must include at least the basic knowledge. $W \subset E$ 
  \item Given that Borhan's objective is to extend Classical Logic and integrate the outcomes of default rules into the base knowledge, it's evident that any extension should inherently encompass the expansion of Classical Logic knowledge. This entails considering the consequences of Classical Logic in both prerequisites and justifications. $Th(E) = E$
  \item A normal extension should also be closed to the default rules. Therefore, as long as there is still a default rule that can be executed, the normal extension has not yet been obtained.
  \item As shown in \cite{reiter1980logic}, the previous properties are not enough to define the normal extension. Each extension must be accompanied by an explanation based on basic knowledge and the rules of descriptive logic and default logic. Consequently, an expansion can only be deemed a normal expansion if it fulfills the three mentioned conditions, in addition to providing a clear explanation for how the extension is achieved based on basic knowledge and rules. This fourth condition is a novel contribution of our work and distinguishes it from previous works.
\end{itemize}
By the mentioned properties and using a rotation process, the process of making extensions can be introduced as follows:
\begin{equation}
E = \bigcup_{i=0}^{\infty}E_{i}
\end{equation}
In the first step, we consider $E_0=Th(W)$. (Some previous works have introduced the first step $E_0=W$, which may produce false results in the prerequisites and justifications of the default rules in the first step \cite{reiter1980logic,sep-logic-nonmonotonic}) Then, using the rule, subsequent expansions are made until a fixed point finally will be reached. \par
\begin{equation}
E_{n+1} = Th(E_n)\cup \{C|\frac{P:J_1, J_2, ..., J_m}{C}\in D, \neg J_i \notin E, P \in E_n\}
\end{equation}
The noticeable and non-monotonic point in this rotation process is that the complement of the justifications must not be in the final normal extension ($E$) and not in the extension of step n ($E_n$). This fact makes it possible that some of the conclusions that are added to the extension will be removed later, and as a result, the conclusions that were only based on them will also be removed. It should be noted that this process is only one of the possible processes to create extensions, and an extension may be obtained without using this method and have the desired properties.

Therefore, if we want to use ($E$) to create any extensions, it cannot be implemented because it was not made in the intermediate steps and we do not have access to it, and if we want to use ($E_n$), we did not logically right process. This problem is seen in the prioritized default logic examples as well. For example, It was observed that generating results with lower priorities in the initial steps could hinder the generation of more reliable results with higher priorities in later steps.
To address this issue, this article presents a solution. First, we suppose that for reaching the extension at step $n$, only the knowledge up to the step before it is necessary. This approach allows for a cyclical process that can identify and manage inconsistencies or violations of justifications. for this purpose, we define $C_n$ as the set of all non-monotonic inference results by the default rules up to the $nth$ step, as follows:

\begin{equation}
\label{eq:cn}
C_{n} = \bigcup_{k=0}^{n} \{C|\frac{P:J_1, J_2, ..., J_m}{C}\in D, \neg J_i \notin E_k, P \in E_K\}
\end{equation}

Also, we define Non-Monotonic Reduced Normal Extension ($\overline{E}_n$) as follows: 

\begin{equation}
\label{eq:enoverbar}
    \overline{E}_n = 
    \begin{cases}
      \forall \delta_i \in D, \forall j \in J_{\delta_i}: \neg j \notin Th\{C_n\}: E_n \\
      \forall \delta_i in D, \exists j \in J_{\delta_i}: \neg j \in Th\{C_n\}: E_n - \bigcup Th\{\neg j\}
    \end{cases}
\end{equation}

Where in equation \ref{eq:enoverbar} $J_\delta$ is considered to be the set of all justifications of rule $\delta$, and $D$ as the set of all default rules, and $Th\{C_n\}$ as the set of all non-monotonic results up to the $nth$ step. Our objective in converting $E_n$ to $\overline{E }_n$ is to delay the creation of non-monotonic results that cause some of the default rules to be non-applicable. Therefore, in the next steps of the expansion, all the results of default rules can be added to the knowledge and the priorities determine the extension status. Without this adjustment, a rule with a lower priority might violate the justification of a higher-priority rule, causing the more valuable rule to lose its applicability. Such a scenario contradicts the meaning of the priority concept. After converting $E_n$ to $\overline{E}_n$, we define the next step of expansion as follows:

\begin{equation}
E_{n+1} = \overline{E}_n \cup \{C|\frac{P:J_1, J_2, ..., J_m}{C}\in D, \neg J_i \notin \overline{E}_n, P \in \overline{E}_n\}
\end{equation}

While the clause $\neg J_i \notin \overline{E}_n$ is not required according to the definition of $\overline{E}_n$, it has been included for the sake of similarity with the standard format of the extension recursive rule.

In this process, all default rules are executed, and the status of these results which were initially discarded due to their presence in the justification of another rule, is determined during the process of the Justification Checking and the Conflict Analyzing. If there are no obstacles to their generation, these results are re-generated by predefined rules and subsequently included in the knowledge during the next expansion stage. Furthermore, for those non-monotonic results that have been reduced and reintroduced by the rules, they do not pose a problem in reaching a fixed point. This is because, before rule execution, they were present in the $nth$ stage, and after rule execution, they will still be present in the $n+1$ stage, even though they were initially reduced and then added again.

The proposed idea is a combination of Reiter's method for calculating the rotation of extensions and Antoniou's method that uses the paths for calculating extensions by forming two sets of $In$ and $Out$. \cite{antoniou1999tutorial, reiter1980logic}.

In the Borhan framework, the basic model that includes classes, relations, individuals, and axioms is considered as $E_0$. Then, by using an incremental descriptive logic reasoner, in the first step, the default rules are executed, and the conclusions are added to the previous knowledge. This process will continue until we finally reach a fixed point. The details of this process will be covered in details in the following section.

\subsection{Variants of Default Logic}
As outlined in \cite{antoniou1999tutorial}, Default Logic serves as a valuable tool for expanding our knowledge when we lack complete information about a problem or situation. Default logic encompasses various variations, including justified, constrained, and prioritized forms. Justified Default Logic indicates that we will stop and accept the current extension if all ways of expanding the extension by applying a new default lead to an inconsistent model. Constrained Default Logic, on the other hand, enforces joint consistency. The primary focus of this article is on Prioritized Default Logic, which considers priorities among defaults, indicating a preference for which default to use in specific situations. It's worth noting that within the Borhan framework, the fundamental components and methods for Default Logic are designed, making it possible to implement other types of Default Logic as well.

\section{System Architecture}
\begin{figure*}[htb!]
\centering
\includegraphics[width=0.95\textwidth,height=5cm]{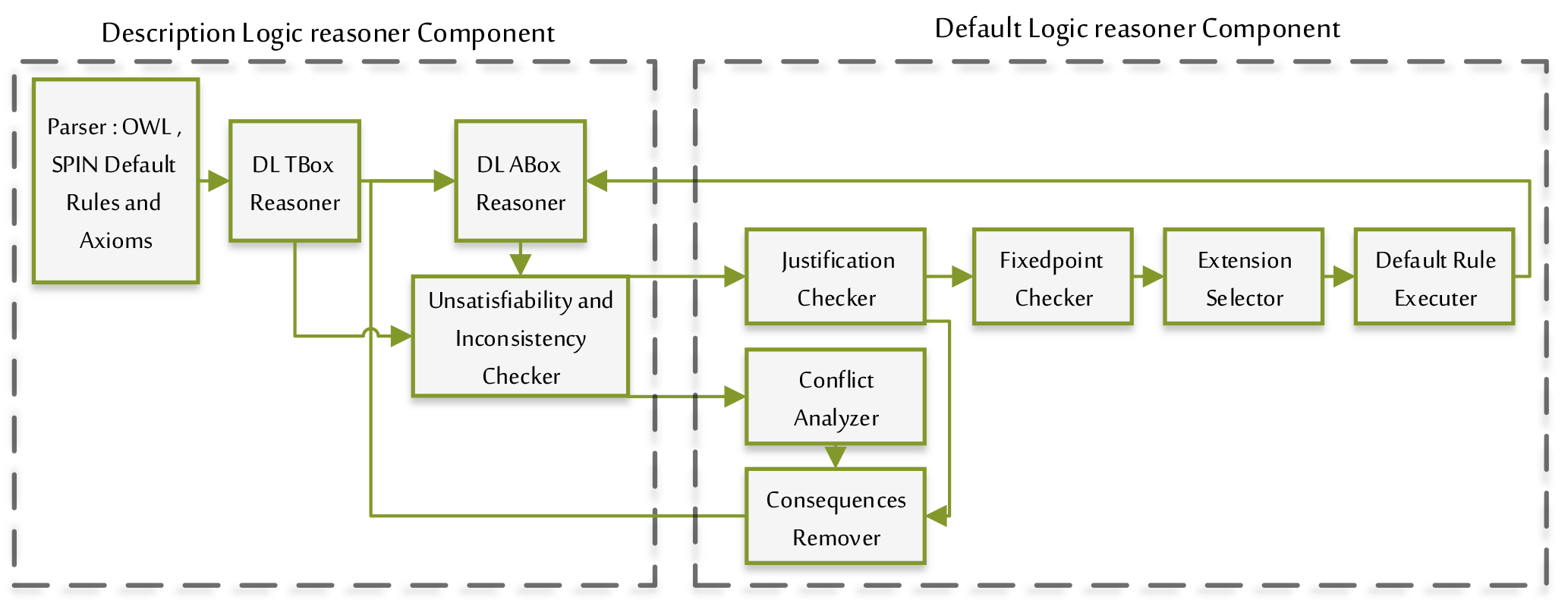}
\caption{The Architecture of Borhan}
\label{architecture}
\end{figure*}

The architecture of Borhan can be seen in Figure \ref{architecture}. As can be seen, Borhan consists of several components that help to solve the problem correctly. In this section, these components will be explained thoroughly. \par
As shown in Figure \ref{architecture}, the Borhan framework is composed of two main parts:
\begin{enumerate}
  \item description logic reasoner
  \item default logic reasoner
\end{enumerate}

The primary objective of a description logic reasoner is to verify consistency, and in case of inconsistency, explain it. Various Descriptive Logic Reasoners, such as Pellet and HermiT, are available for this purpose. However, we opted to employ the Borhan Incremental Descriptive Logic Reasoner due to the frequent need for explanations in default logic reasoning. Additionally, incremental addition of knowledge is a common aspect in non-monotonic reasoning, which the Borhan reasoner can effectively handle.

\begin{algorithm}
\caption{Algorithm of the Borhan framework}
\label{borh_alg}
\textbf{Input: } basic model ($E_0$) includes classes, relations, individuals, and axioms; default\_rules. \par
\begin{algorithmic}[1]
\State $E_n = E_0$
\State tbox\_discrtiptive\_reasoning($E_n$)
\While {true}
    \State abox\_discrtiptive\_reasoning($E_n$)
    \State conflict, reason\_conflict = Inconsistency\_checker($E_n$) 
    \If {conflict == true}
        \If {reason\_conflict == default}
            \State cause\_conflict = conflict\_analyzer($E_n$) \Comment{cause of the conflict with the minimum default value}
            \State consequence\_remover($E_n$, cause\_conflict) \Comment{remove the given axiom and all its consequences}
        \Else
            \State \Return unresolvable
        \EndIf
    \Else
        \State justification, cause\_justification = justification\_checker($E_n$) 
        \If {justification == true}
            \State consequence\_remover($E_n$, cause\_justification)
        \Else 
            \State fixed\_point = fixed\_point\_checker($E_n$)
            \If {fixed\_point == true}
                \State \Return $E_n$
            \EndIf
            \State $E_n$ = extension\_selector($E_n$, default\_rules)
            \State $E_n$ = default\_rule\_executor($E_n$, default\_rules)
        \EndIf
    \EndIf
\EndWhile

\end{algorithmic}
\end{algorithm}

\subsection{Conflict Analyzer}
\label{sec:conf_analyzer}
The inconsistency checker module in the description logic reasoner has two possible outputs. When conflicts exist in the knowledge graph due to default rules, the conflict analyzer is invoked. The purpose of this module is simple: it identifies the cause of the conflict with the lowest default value, or in other words, the minimum priority. For instance, in the penguin problem, there is a conflict between being non-flying, a characteristic of penguins, and being flying, a characteristic of birds, which penguins are a subset of. In this scenario, being flying has a lower priority than being non-flying. Therefore, the output of this module would designate "being a flying" as the minimum value for the conflict.

\begin{figure*}[htb!]
\centering
\includegraphics[width=0.5\textwidth,height=6cm]{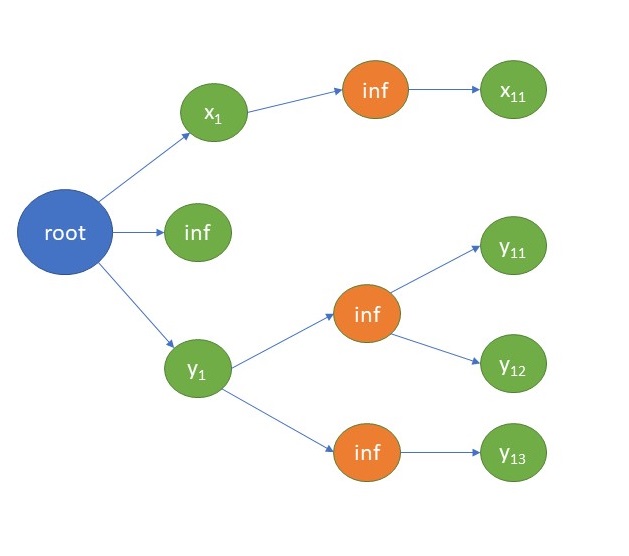}
\caption{An example of the conflict tree as the input of the conflict analyzer.}
\label{fig:conflict_tree}
\end{figure*}

The conflict analyzer's input is structured as a tree. The tree's root node marks the conflict's initiation, while the other nodes in the tree can be either statements or explanation sets. The initial level of nodes within the tree consists of statements that form the basis for the conflict. Each of these nodes is assigned a priority value, determined as the minimum value among the rule order values capable of concluding that specific statement. If a statement is asserted, its node value is set to infinity. For nodes that are inferred and not asserted, there are child nodes representing explanation sets. Each explanation set node serves as the parent for a group of statements that, when taken together, can lead to the conclusion of the parent statement node within the explanation set. The priority value for each explanation set node is set to infinity. The children of an explanation set node can either be other inferred or asserted statements. In Figure \ref{fig:conflict_tree}, we've provided an example of a conflict tree. Within this figure, statement nodes are depicted with green circles, and explanation nodes are represented by orange circles. Notably, the green nodes can have either an infinity value, signifying assertion, or a typical value, denoting inference and a priority value. All the orange circles have infinite values, as they are indicative of explanation set nodes. The statements at the third depth level are assumed to all have statements as their children with infinite values and are not shown to maintain simplicity.

To identify the node with the lowest priority in the conflict tree whose removal resolves the conflict, we've developed an algorithm based on the Min-Max approach. This choice is motivated by the need to select the minimum priority when considering sibling statement nodes (indicated as green nodes in Figure \ref{fig:conflict_tree}) and the maximum priority when dealing with sibling explanation set nodes (indicated as orange nodes in Figure \ref{fig:conflict_tree}).

The implemented algorithm begins at the root of the tree, where each parent node initiates recursive calls to its child nodes for priority calculation. For siblings sharing the same parent, the algorithm determines the following: a) if the siblings are statement nodes, the parent receives the minimum priority value among them. b) if the siblings are explanation set nodes, the parent receives the maximum priority value. Upon receiving priority values from its children, each parent node conducts the following operation: a) a comparison is made between the priority value of the parent and the value returned by its children. b) the minimum of these values is selected to determine the final priority value for the parent node. Ultimately, this process continues up the tree until the root node determines which node should be removed to resolve the conflict.

\subsection{Justification Checker}
The Justification Checker module serves as an alternative output for the inconsistency checker and is invoked when there is no inconsistency in the knowledge graph. This module's purpose is to examine whether the complement of the justification for each axiom is present in the knowledge graph. In default logic, the complement of a justification for an axiom may be inferred later in the reasoning process. If the complement of a justification for an axiom is found in the knowledge graph, the axiom should be eliminated. For instance, if the justification for playing football is a "not raining day," and an axiom in the knowledge graph asserts the presence of a "rainy day," the playing football axiom should be removed.

\subsection{Consequence Remover}
If either the conflict analyzer or the justification checker determines that an axiom needs to be removed, that axiom is passed to the consequence remover module. The function of this module is to identify all consequences which haven't any other explanations associated with the given axiom and eliminate both the axiom itself and its related consequences.

\subsection{Fixed point Checker}
Towards the conclusion of this process, there is a module called the Fixedpoint Checker, which assesses whether a fixed point has been reached. As explained in the prior section, a fixed point is a consistent state where no applicable default rules remain.

\subsection{Extension Selector}
As previously stated, the conversion of $E_n$ to $\overline{E}_n$ was done to achieve a well-defined recursive rule. In the implementation section, this transformation is also executed using SPARQL commands. Given that the default rules are expressed using SPINs, this process is straightforward.

\subsection{Default Rule Executor}
Because SPIN is employed as the rule-writing language, the rule executor essentially functions as a SPARQL execution engine. It's important to note that, for this non-monotonic reasoning system's classical logic reasoner, we utilized the Borhan incremental reasoner. Consequently, it is imperative that, alongside the implementation of default rules, all results, along with their prerequisites, justifications, and orders, are provided in a standardized format to the incremental reasoner. This ensures that the process of knowledge development or correction, such as in the Consequence Remover section, is executed accurately.

\begin{table}[]
\caption{Collection of the Experimental Set}
\label{table:experiment_set}
\begin{adjustbox}{width=1.0\textwidth, height=7cm}
\begin{tabular}{|l|l|l|l|l|l|}
\hline
\# & \textbf{Title}              & \textbf{Textual Form}                                                                                             & \textbf{Symbolic Form}                                                                                                                                      & \textbf{Input Data}                                                                                             & \textbf{Expected Output Data}                                                                                                           \\ \hline
1  & Basic Default Rule \cite{reiter1980logic} & \begin{tabular}[c]{@{}l@{}}Birds usually fly.\\ Penguins are birds. \\ Penguins do not fly.\end{tabular} & \begin{tabular}[c]{@{}l@{}}$\frac{Bird(x):Flying(x)}{Flying(x)}$\\ $Penguin(x) \supset Bird(x)$\\ $Penguin(x) \supset \neg Flying(x)$\end{tabular} & \begin{tabular}[c]{@{}l@{}}$Penguin(tweety)$\\ \\ ----------------------\\ $Bied(tweety)$\end{tabular} & \begin{tabular}[c]{@{}l@{}}$Bird(tweety)$\\ $\neg Flying(tweety)$\\ --------------------------\\ $Flying(tweety)$\end{tabular} \\ \hline
 2  & Transitivity in Default rule \cite{antoniou1999tutorial}               &      \begin{tabular}[c] {@{}l@{}} Usually, my friends’ friends \\ are also my friends. \end{tabular}                                                                                     &      $\frac{friend(X, Y) \land friend(Y, Z): friend(X, Z)}{friend(X, Z)}$                                                                                                                                                    &   \begin{tabular}[c] {@{}l@{}}    friend(tom,bob) \\ friend(bob,sally) \\ friend(sally,tina)     \end{tabular}                                                                                          &       friend(tom,tina)                                                                                                                         \\ \hline
 3  &     Nixon \cite{bozzato2022reasoning}             &       \begin{tabular}[c] {@{}l@{}}   Usually, every Quacker is a Pacifist \\ Usually, every Republican is a non-Pacifist  \end{tabular}                                                                                               &      \begin{tabular}[c] {@{}l@{}}       $\frac{Quacker(x): Pacifist(x)}{Pacifist(x)}$ \\ $\frac{Republican(x): \neg Pacifist(x)}{\neg Pacifist(x)}$  \end{tabular}                                                                                                                                       &     \begin{tabular}[c] {@{}l@{}}     Quacker (Nixon) \\ Republican (Nixon)
   \end{tabular}                                                                                                &    \begin{tabular}[c] {@{}l@{}}    two extensions: \\ $Pacifist(Nixon)$ \\ $\neg Pacifist(Nixon)$    \end{tabular}                                                                                                                    \\ \hline 4 & Self-defeater \cite{sep-logic-nonmonotonic} & \begin{tabular}[c] {@{}l@{}} in the absence of any \\ information to the contrary that  \\ x is not Thing, it inferred that x  \\ is Thing \end{tabular} & $\frac{True: \neg Man(x)}{Man(x)}$ & $Man(x)$ & \begin{tabular}[c] {@{}l@{}} Nothing  \\ The reasoner must not\\ fall in the loop \end{tabular}\\ \hline
   5 & Bird or Dog \cite{brewka2008nonmonotonic} & \begin{tabular}[c] {@{}l@{}} Usually, if an animal sings, it is a bird. \\ Birds and dogs are pets \\ Dogs are not birds and vice versa. \\ Usually, pets are dogs. \end{tabular} & \begin{tabular}[c] {@{}l@{}} $\frac{Animal(x) \land Sings(x): Bird(x)}{Bird(x)}$ \\ $Bird(x) \lor Dog(x) \supset Pet(x)$ \\ $Bird(x) \supset \neg Dog(x)$  \\ $\frac{Pet(x): Dog(x)}{Dog(x)}$ \end{tabular}& \begin{tabular}[c] {@{}l@{}} $Animal(a)$ \\ $Sings(a)$ \end{tabular} & \begin{tabular}[c] {@{}l@{}} $Bird(a)$ \\ $Pet(a)$ \\ $\neg Dog(a)$ \end{tabular}\\ \hline 
   6 & Dropouts and employed \cite{antoniou1999tutorial} & \begin{tabular}[c] {@{}l@{}}  Typically, high school dropouts are adults. \\ Typically, adults who aren’t dropouts\\ are employed.
 \end{tabular} & \begin{tabular}[c] {@{}l@{}} $\frac{dropout(x):adult(x)}{adult(x)}$ \\ $\frac{adult(x):employed(x) \land \neg dropout(x)}{employed(x)}$ \\ --------------------------------------- \\ $\delta_1 = \frac{dropout(x):adult(x)}{adult(x)}$ \\ $\delta_2 = \frac{adult(x):employed(x)}{employed(x)}$ \\ $\delta_3 = \frac{dropout(x):\neg employed(x)}{\neg employed(x)}$ \\ $\delta_3 \ll \delta_2 \ll \delta_1 $ \end{tabular} & Bill is a high school dropout. & Bill is an adult. \\ \hline 7 & Broken-arms \cite{brewka2008nonmonotonic}  &  \begin{tabular}[c] {@{}l@{}} If the unbroken and usable arm \\ is consistent with our \\ knowledge, then by default, we \\ assume it is usable. \end{tabular} & $\frac{True:Useable(x) \land \neg Broken(x)}{Useable(x)}$ & \begin{tabular}[c] {@{}l@{}} $Broken(Right) \lor $ \\ $Broken(Left)$ \end{tabular} & \begin{tabular}[c] {@{}l@{}} it has two extensions: \\ $Useable(Right)$ \\  $Useable(Left)$ \end{tabular} \\ \hline 8 & Abnormal default rules \cite{antoniou1999tutorial} & \begin{tabular}[c] {@{}l@{}} As long as it doesn't snow, \\ the football match will be played. \end{tabular} & \begin{tabular}[c] {@{}l@{}} $\delta_1 = \frac{Football(y,x): \neg Snow(x)}{TakesPlace(y)}$ \\ $\delta_2 = \frac{Dark(x):\neg Sunny(x)}{\neg Sunny(x)}$ \\ $\delta_3 = \frac{\neg Sunny(x):Snow(x))}{Snow(x))}$ \end{tabular} & \begin{tabular}[c] {@{}l@{}} $Football(a,b)$ \\ $Dark(b)$ \end{tabular} & \begin{tabular}[c] {@{}l@{}}  $\neg Sunny(b)$ \\ $Snow(b)$ \end{tabular} \\ \hline 9 & Priorities between Defaults \cite{lifschitz1989benchmark} & \begin{tabular}[c] {@{}l@{}} If there is a contradiction \\ between Mary's and Jack's statements, \\ Mary's statement is more reliable. \end{tabular} & \begin{tabular}[c] {@{}l@{}}  $\delta_1 = \frac{JackAsserted(x):Right(x)}{Right(x)}$ \\ $\delta_2 = \frac{MaryAsserted(x):Right(x)}{Right(x)}$ \\  
  $InContradict(x,y) \land Right(y) \supset \neg Right(x)$ \\ $\delta_2 \ll \delta_1$ \end{tabular} & \begin{tabular}[c] {@{}l@{}} $JackAsserted(S_1)$ \\ $MaryAsserted(S_2)$ \\ $InContradict(S_1,S_2)$ \end{tabular} & $Right(S_2)$ \\ \hline 10 & Oviparous vs Mammal (Oviparous)& \begin{tabular}[c] {@{}l@{}} Penguins are usually birds. \\ Birds are usually oviparous. \\ Penguins are usually in Antarctica. \\ Every animal in Antarctica is \\ usually a mammal. \\ Mammals are not usually oviparous. \end{tabular} & \begin{tabular}[c] {@{}l@{}} $\delta_1 = \frac{Penguin(x):Bird(x)}{Bird(x)}$ \\ $\delta_2 = \frac{Bird(x):Oviparous(x)}{Oviparous(x)}$ \\ $\delta_3 = \frac{Penguin(x): InAntarctica(x)}{InAntarctica(x)}$ \\ $\delta_4 = \frac{InAntarctica(x):Mammal(x)}{Mammal(x)}$ \\ $\delta_5 = \frac{Mammal(x): \neg Oviparous(x)}{\neg Oviparous(x)}$ \\ $\delta_6 = \frac{Oviparous(x): \neg Mammal(x)}{\neg Mammal(x)} $ \\ $\delta_1 \ll \delta_6 \ll \delta_5, \delta_2 \ll \delta_3 \ll \delta_4 $ \end{tabular} & Penguin(a) & \begin{tabular}[c] {@{}l@{}} Bird(a) \\ Oviparous(a) \\ InAntarctica(a) \end{tabular} \\ \hline  11 & Oviparous vs Mammal (Mammal)& \begin{tabular}[c] {@{}l@{}} Penguins are usually birds. \\ Birds are usually oviparous. \\ Penguins are usually in Antarctica. \\ Every animal in Antarctica is \\ usually a mammal. \\ Mammals are not usually oviparous. \end{tabular} & \begin{tabular}[c] {@{}l@{}} $\delta_1 = \frac{Penguin(x):Bird(x)}{Bird(x)}$ \\ $\delta_2 = \frac{Bird(x):Oviparous(x)}{Oviparous(x)}$ \\ $\delta_3 = \frac{Penguin(x): InAntarctica(x)}{InAntarctica(x)}$ \\ $\delta_4 = \frac{InAntarctica(x):Mammal(x)}{Mammal(x)}$ \\ $\delta_5 = \frac{Mammal(x): \neg Oviparous(x)}{\neg Oviparous(x)}$ \\ $\delta_6 = \frac{Oviparous(x): \neg Mammal(x)}{\neg Mammal(x)} $ \\ $\delta_1 \ll \delta_5 \ll \delta_3 \ll \delta_4 \ll \delta_2 \ll \delta_6 $ \end{tabular} & Penguin(a) & \begin{tabular}[c] {@{}l@{}} Bird(a) \\ $\neg Oviparous(a)$ \\ InAntarctica(a) \\ Mammal(a) \end{tabular} \\ \hline
\end{tabular}
\end{adjustbox}
\end{table}

\section{Experiments}
To demonstrate the efficacy of our model, we assembled an experimental set from default logic papers, encompassing various default examples that our model can effectively manage and solve. This experimental dataset is shown in Table \ref{table:experiment_set}, which contains a variety of examples featuring distinct titles, each representing a distinct problem. For each example, both the textual and symbolic representations are provided, along with corresponding input and expected output data.

Prior logical systems, such as Prolog and Datalog, are capable of managing basic non-monotonic reasoning, as exemplified by example 1 in our experimental dataset. In this type of example, a rule yields a result if and only if the prerequisites of the rule are present in the knowledge base, and the negation of justifications is absent. Example 1 showcases the well-known penguin problem, in which we deduce that a bird can fly unless it happens to be a penguin. Nonetheless, previous systems are incapable when it comes to addressing prioritized non-monotonic problems, as exemplified by the example in the 6th, 8th, 9th, 10th, and 11th rows of the table. Notably, these situations often lead to inconsistencies between two axioms, with one of them derived from a higher-order rule. The Borhan framework proves its proficiency in handling such examples. In these cases, the conflict analyzer module, previously described in Section \ref{sec:conf_analyzer}, identifies the axiom with a lower order, which will be removed later. when orders are equal between several axioms, the component will provide all of them, allowing the user to decide which should be removed. Thus, our model demonstrates its capacity to effectively tackle examples like the Nixon example in the 3rd row of the table.

The Oviparous vs. Mammals problems which were found in the 10th and 11th rows of our experimental set, illustrate the novelty of our model. The key distinction between these two instances lies in the priority of rule $\delta_2$. In the 10th example, it is equivalent to $\delta_5$ and higher than $\delta_4$, while in the 11th row, it assumes the minimum value. In both of these examples, there is an inconsistency between being oviparous and being a mammal. The presence of mammal attributes leads to non-oviparous characteristics, representing a conflict between rules $\delta_2$ and $\delta_4$. In both scenarios, the conditions of being oviparous and being a mammal are addressed in the second run, utilizing the information related to being a bird and residing in Antarctica. Being oviparous hinders the execution of rule $\delta_5$, as it is a part of the rule's justification. Since "mammal" is a prerequisite for this rule, it should be removed.

In the 10th example, the $\delta_2$ rule takes precedence over the $\delta_4$ rule, making the removal of the "mammal" attribute the correct course of action. However, in the 11th example, the $\delta_2$ rule carries the lowest priority, and removing the "mammal" attribute due to the presence of "oviparous" is not a valid action solely based on the fact that it was created first. To address this issue, we propose the removal of the "oviparous" axiom and rely on rule priorities to determine which axiom to retain and which to discard. After removing the "oviparous" axiom, in the 10th example, the $\delta_2$ rule, having higher priority, thus makes the "oviparous" axiom again. In contrast, in the 11th example, where this rule has lower priority, it is not executed again, resulting in "mammal" and "not being oviparous" as the outcomes. This innovative approach to adjusting the process of creating extensions to resolve the mentioned problem is a novelty of our article.

\section{Conclusion}

In this article, we introduce the Borhan framework, a logical system specifically designed to address prioritized default logic problems. We have introduced Non-Monotonic Reduced Normal Extension and a novel idea for the definition of extensions in default logic, enhancing its solvability for certain problems. Our model comprises multiple integral components, each playing a crucial role in problem resolution. Significantly, our system bridges the gap between theoretical default logic and a practical, implementable solution suitable for various problem domains. Given its completely logical and explainable nature, our system is well-suited for high-risk applications such as legal judgment and medical diagnosis. It represents the pioneering effort in this field and offers the potential for expansion to tackle different variants of default logic. Furthermore, the adaptability of our model allows for making a model for other logical problems, such as ramification problems, through integration with dynamic systems. Additionally, combining our model with neural networks can enhance decision-making efficiency and speed.

%Bibliography
\bibliographystyle{unsrt}  
\bibliography{references}

\begin{thebibliography}{10}

\bibitem{sep-logic-nonmonotonic}
Christian Strasser and G.~Aldo Antonelli.
\newblock {Non-monotonic Logic}.
\newblock In Edward~N. Zalta, editor, {\em The {Stanford} Encyclopedia of Philosophy}. Metaphysics Research Lab, Stanford University, {S}ummer 2019 edition, 2019.

\bibitem{mcdermott1980non}
Drew McDermott and Jon Doyle.
\newblock Non-monotonic logic i.
\newblock {\em Artificial intelligence}, 13(1-2):41--72, 1980.

\bibitem{reiter1980logic}
Raymond Reiter.
\newblock A logic for default reasoning.
\newblock {\em Artificial intelligence}, 13(1-2):81--132, 1980.

\bibitem{garcia2004defeasible}
Alejandro~J Garc{\'\i}a and Guillermo~R Simari.
\newblock Defeasible logic programming: An argumentative approach.
\newblock {\em Theory and practice of logic programming}, 4(1-2):95--138, 2004.

\bibitem{bozzato2022reasoning}
Loris Bozzato, Thomas Eiter, and Luciano Serafini.
\newblock Reasoning on with defeasibility in asp.
\newblock {\em Theory and Practice of Logic Programming}, 22(2):254--304, 2022.

\bibitem{brewka1994reasoning}
Gerhard Brewka.
\newblock Reasoning about priorities in default logic.
\newblock In {\em AAAI}, volume 1994, pages 940--945, 1994.

\bibitem{rudinger2020thinking}
Rachel Rudinger, Vered Shwartz, Jena~D Hwang, Chandra Bhagavatula, Maxwell Forbes, Ronan Le~Bras, Noah~A Smith, and Yejin Choi.
\newblock Thinking like a skeptic: Defeasible inference in natural language.
\newblock In {\em Findings of the Association for Computational Linguistics: EMNLP 2020}, pages 4661--4675, 2020.

\bibitem{arieli2023postulate}
Ofer Arieli, AnneMarie Borg, and Christian Stra{\ss}er.
\newblock A postulate-deriven study of logical argumentation.
\newblock {\em Artificial Intelligence}, page 103966, 2023.

\bibitem{chen1996extending}
Weidong Chen.
\newblock Extending prolog with nonmonotonic reasoning.
\newblock {\em The Journal of logic programming}, 27(2):169--183, 1996.

\bibitem{leone2006dlv}
Nicola Leone, Gerald Pfeifer, Wolfgang Faber, Thomas Eiter, Georg Gottlob, Simona Perri, and Francesco Scarcello.
\newblock The dlv system for knowledge representation and reasoning.
\newblock {\em ACM Transactions on Computational Logic (TOCL)}, 7(3):499--562, 2006.

\bibitem{ricca2007disjunctive}
Francesco Ricca and Nicola Leone.
\newblock Disjunctive logic programming with types and objects: The dlv+ system.
\newblock {\em Journal of Applied Logic}, 5(3):545--573, 2007.

\bibitem{cholewinski1996default}
Pawel Cholewinski, V~Wiktor Marek, and Miroslaw Truszczynski.
\newblock Default reasoning system deres.
\newblock {\em KR}, 96:518--528, 1996.

\bibitem{chen2010dl2asp}
Yin Chen, Hai Wan, Yan Zhang, and Yi~Zhou.
\newblock dl2asp: implementing default logic via answer set programming.
\newblock In {\em European Workshop on Logics in Artificial Intelligence}, pages 104--116. Springer, 2010.

\bibitem{brewka2002logic}
Gerhard Brewka.
\newblock Logic programming with ordered disjunction.
\newblock In {\em AAAI/IAAI}, pages 100--105, 2002.

\bibitem{devlin2018bert}
Jacob Devlin, Ming-Wei Chang, Kenton Lee, and Kristina Toutanova.
\newblock Bert: Pre-training of deep bidirectional transformers for language understanding.
\newblock {\em arXiv preprint arXiv:1810.04805}, 2018.

\bibitem{yu2023nature}
Fei Yu, Hongbo Zhang, and Benyou Wang.
\newblock Nature language reasoning, a survey.
\newblock {\em arXiv preprint arXiv:2303.14725}, 2023.

\bibitem{liu2019roberta}
Yinhan Liu, Myle Ott, Naman Goyal, Jingfei Du, Mandar Joshi, Danqi Chen, Omer Levy, Mike Lewis, Luke Zettlemoyer, and Veselin Stoyanov.
\newblock Roberta: A robustly optimized bert pretraining approach.
\newblock {\em arXiv preprint arXiv:1907.11692}, 2019.

\bibitem{raffel2020exploring}
Colin Raffel, Noam Shazeer, Adam Roberts, Katherine Lee, Sharan Narang, Michael Matena, Yanqi Zhou, Wei Li, and Peter~J Liu.
\newblock Exploring the limits of transfer learning with a unified text-to-text transformer.
\newblock {\em The Journal of Machine Learning Research}, 21(1):5485--5551, 2020.

\bibitem{lewis2019bart}
Mike Lewis, Yinhan Liu, Naman Goyal, Marjan Ghazvininejad, Abdelrahman Mohamed, Omer Levy, Ves Stoyanov, and Luke Zettlemoyer.
\newblock Bart: Denoising sequence-to-sequence pre-training for natural language generation, translation, and comprehension.
\newblock {\em arXiv preprint arXiv:1910.13461}, 2019.

\bibitem{radford2019language}
Alec Radford, Jeffrey Wu, Rewon Child, David Luan, Dario Amodei, Ilya Sutskever, et~al.
\newblock Language models are unsupervised multitask learners.
\newblock {\em OpenAI blog}, 1(8):9, 2019.

\bibitem{baader2003description}
Franz Baader, Diego Calvanese, Deborah McGuinness, Peter Patel-Schneider, Daniele Nardi, et~al.
\newblock {\em The description logic handbook: Theory, implementation and applications}.
\newblock Cambridge university press, 2003.

\bibitem{baader1995embedding}
Franz Baader and Bernhard Hollunder.
\newblock Embedding defaults into terminological knowledge representation formalisms.
\newblock {\em Journal of Automated Reasoning}, 14(1):149--180, 1995.

\bibitem{antoniou1999tutorial}
Grigoris Antoniou.
\newblock A tutorial on default logics.
\newblock {\em ACM Computing Surveys (CSUR)}, 31(4):337--359, 1999.

\bibitem{brewka2008nonmonotonic}
Gerhard Brewka, Ilkka Niemel{\"a}, and Miros{\l}aw Truszczy{\'n}ski.
\newblock Nonmonotonic reasoning.
\newblock {\em Foundations of Artificial Intelligence}, 3:239--284, 2008.

\bibitem{lifschitz1989benchmark}
Vladimir Lifschitz.
\newblock Benchmark problems for formal nonmonotonic reasoning: Version 2.00.
\newblock In {\em Non-Monotonic Reasoning: 2nd International Workshop Grassau, FRG, June 13--15, 1988 Proceedings 2}, pages 202--219. Springer, 1989.

\end{thebibliography}

\end{document}